\documentclass{article}

\usepackage[numbers]{natbib}         
\usepackage[colorlinks]{hyperref}    
\usepackage[english]{babel} 
\usepackage{amssymb}
\usepackage{amsmath}
\usepackage{txfonts}
\usepackage{mathdots}
\usepackage[classicReIm]{kpfonts}
\usepackage[dvips]{graphicx} 
\usepackage[a4paper, portrait, margin=1in]{geometry}
\usepackage{tabularx}
\usepackage{longtable}
\usepackage{multirow}
\usepackage{booktabs}
\usepackage[labelsep=period]{caption}
\usepackage{makecell}
\begin{document}
	\setlength{\parindent}{0pt}
	\setlength{\parskip}{1ex}
	
	\textbf{\Large Hippocampus Substructure Segmentation Using Morphological Vision Transformer Learning}
	
	\bigbreak

	Yang Lei$^{1}$, Yifu Ding$^{1}$, Richard L.J. Qiu$^{1}$, Tonghe Wang$^{2}$, Justin Roper$^{1}$, Yabo Fu$^{2}$, Hui-Kuo Shu$^{1}$, 
	Hui Mao$^{3}$ and Xiaofeng Yang$^{1*}$

	1Department of Radiation Oncology and Winship Cancer Institute, Emory University, Atlanta, GA 30308
	
	2Department of Medical Physics, Memorial Sloan Kettering Cancer Center, New York, NY, 10065
	
	3Department of Radiology and Imaging Sciences and Winship Cancer Institute, Atlanta, GA 30308

	\bigbreak
	\bigbreak
	\bigbreak

	\textbf{*Corresponding author: }
	
	Xiaofeng Yang, PhD
	
	Department of Radiation Oncology
	
	Emory University School of Medicine
	
	1365 Clifton Road NE
	
	Atlanta, GA 30322
	
	E-mail: xiaofeng.yang@emory.edu

	\bigbreak
	\bigbreak
	\bigbreak
	\bigbreak
	\bigbreak
	\bigbreak

	\textbf{Abstract}

	Background: The hippocampus plays a crucial role in memory and cognition. Because of the associated toxicity from whole brain radiotherapy, more advanced treatment planning techniques prioritize hippocampal avoidance, which depends on an accurate segmentation of the small and complexly shaped hippocampus. 
	Purpose: To achieve accurate segmentation of the anterior and posterior regions of the hippocampus from T1 weighted (T1w) MRI images, we developed a novel model, Hippo-Net, which uses a mutually enhanced strategy.
	
	Methods: The proposed model consists of two major parts: 1) a localization model is used to detect the volume-of-interest (VOI) of hippocampus. 2) An end-to-end morphological vision transformer network is used to perform substructures segmentation within the hippocampus VOI. The substructures include the anterior and posterior regions of the hippocampus, which are defined as the hippocampus proper and parts of the subiculum. The vision transformer incorporates the dominant features extracted from MRI images, which are further improved by learning-based morphological operators. The integration of these morphological operators into the vision transformer increases the accuracy and ability to separate hippocampus structure into its two distinct substructures.
	
	A total of 260 T1w MRI datasets from Medical Segmentation Decathlon dataset were used in this study. We conducted a five-fold cross-validation on the first 200 T1w MR images and then performed a hold-out test on the remaining 60 T1w MR images with the model trained on the first 200 images. The segmentations were evaluated with two indicators, 1) multiple metrics including the Dice similarity coefficient (DSC), 95th percentile Hausdorff distance (HD95), mean surface distance (MSD), volume difference (VD) and center-of-mass distance (COMD); 2) Volumetric Pearson correlation analysis.
	
	Results: In five-fold cross-validation, the DSCs were 0.900±0.029 and 0.886±0.031for the hippocampus proper and parts of the subiculum, respectively. The MSD were 0.426±0.115mm and 0.401±0.100 mm for the hippocampus proper and parts of the subiculum, respectively.
	Conclusions: The proposed method showed great promise in automatically delineating hippocampus substructures on T1w MRI images. It may facilitate the current clinical workflow and reduce the physicians’ effort.
	
	
	\bigbreak
	\bigbreak

	\noindent 
	\section{ INTRODUCTION}
	
	The hippocampus is a pair of medial and subcortical brain structures located in proximity to the temporal horn of the lateral ventricles, which is an active research area due to its implication in memory and neuropsychiatric disorders.\cite{RN5} In radiation therapy, hippocampal avoidance whole brain radiation using volumetric modulated arc therapy (VMAT) plus the medication memantine has been shown to preserve cognitive function without compromising progression-free survival or overall survival when compared to classic whole brain radiation therapy plus memantine.\cite{RN22, RN3} In Alzheimer’s Disease (AD), the progression of AD occurs from the trans-entorhinal cortex to the hippocampus, and finally to the neocortex.\cite{RN3} These progression steps depend on the severity of the neurofibrillary tangles found in neuropathological studies. However, similar patterns can also be observed in the progress of brain atrophy found on MRI imaging studies. The atrophy of hippocampus measured from MRIs can be used as an early sign of AD progression.\cite{RN4} Additionally, evidence of hippocampal atrophy as measured from MRIs can occur before the onset of clinical symptoms.\cite{RN5} Therefore, accurate segmentation of the hippocampus from MRIs is a meaningful task in medical image analysis across multiple disciplines.\cite{RN12} 
	
	To determine whether the hippocampus is atrophic, clinicians often need to segment the bilateral hippocampus on MRI scans and analyze their shape and volume.\cite{RN14, RN13} This task is difficult, however, due to several factors. Firstly, the hippocampus has low contrast with the surrounding tissues on MRI scans,\cite{RN15} since it is a gray matter structure. Secondly, the hippocampus has an irregular shape leading to a blurred boundary in cross-sectional slices.\cite{RN16} Thirdly, the hippocampus is a small structure with limited volume as compared to other structures that are routinely delineated as organs-at-risk (OARs) in radiation therapy.\cite{RN17} Finally, there are large variations in the size and shape of the hippocampus across patients.\cite{RN18} Therefore, accurate and automatic segmentation of hippocampus is a challenging task. Until now, manual segmentation of hippocampus is still the standard in clinical practice.\cite{RN19} However, manual segmentation is a tedious and error-prone process, which limits its application in big data and clinical practice. Thus, many efforts have been devoted to developing computer-aided diagnostic systems for automated segmentation of the hippocampus.
	
	The existing automatic hippocampal segmentation methods can be categorized into two main types: atlas-based methods and machine learning-based methods. Atlas-based methods can be further divided based on the number of atlases used in the segmentation process into single-atlas-based, average-shape atlas-based, and multi-atlas-based approaches. For instance, Haller et al. first proposed to use the single-atlas-based approach for hippocampal segmentation.\cite{RN7, RN6} However, single-atlas-based approaches are limited by inter-patient variations. To address this, average shape-based mapping approaches were proposed to overcome such limitations, but the segmentation results depend on the alignment quality of the target and average maps. Thus, a priori knowledge of medical mapping was incorporated into the multi-atlas-based segmentation approach. For example, Wang et al. proposed a robust discriminative multi-atlas label fusion approach to segment hippocampus by building the conditional random field (CRF) model that combines distance metric learning and graph cuts.\cite{RN8} Wang’s approach is a patch embedding multi-atlas label fusion method that utilizes only the relationship between the target block and the atlas block, and ignores the possibility that unrelated atlas blocks may dominate the voting process. Existing atlas-based methods do not consider the anatomical differences in hippocampus among patients, and do not consider the correlation between atlases.
	 
	Machine learning-based methods can be further classified into traditional machine learning-based approaches and deep learning-based approaches. Traditional machine learning-based approaches mainly include support vector machine (SVM), Markov random field (MRF), principal component analysis (PCA), et al.\cite{RN28, RN181} For instance, Hao et al. proposed a local label learning strategy to estimate segmentation labels of target images by using SVM with image intensity and texture features.\cite{RN20} However, these traditional approaches to machine learning rely heavily on the quality of handcrafted features, and further suffer from slow segmentation, susceptibility to noise interference, and insufficient generalization performance.\cite{RN128} 
	
	Because convolutional neural network (CNN) models can automatically extract the pixel feature information from images, they have been widely used in multiple medical image analysis tasks.\cite{RN185} For example, CNN-based models can be used to segment the hippocampus from MRIs.\cite{RN21} Qiu et al. proposed a multitask 3D U-net framework for hippocampus segmentation by minimizing the difference between the targeted binary mask and the model prediction, and optimizing an auxiliary edge-prediction task.\cite{RN9} Cao et al. developed a two-stage segmentation method to perform the task of 3D hippocampus segmentation by localizing multi-size candidate regions and fusing the multi-size candidate regions.\cite{RN10} These methods show promising results, demonstrating the potential of CNN-based models to improve the efficiency and accuracy of hippocampus segmentation.
	
	However, most existing deep learning-based methods ignore the spatial information of the hippocampus relative to the entirety of the human brain. As a result, they cannot effectively fuse the shape features and the semantic features, which leads to lower segmentation accuracy. Hippocampal tracing began from anterior where the head is visible as an enclosed gray matter structure inferior to the amygdala, and continued posteriorly using surrounding white matter or CSF as boundaries. Subiculum (posterior parts of hippocampus) was included in the hippocampus. Delineation stopped when the wall of the ventricle was visibly contiguous with the fimbria. The subiculum occupies a portion of the para-hippocampal gyrus in the mesial temporal lobe and is a component of the medial temporal memory system. Therefore, in this work, we aim to develop a novel deep network framework to segment the hippocampus by introducing a spatial attention mechanism to capture the spatial location information of the hippocampus relative to the brain. We also designed a cross-layer dual encoding shared decoding network to extract the semantic characteristics of the hippocampus. By combining the spatial location information and semantic characteristics of the hippocampus, we enhanced the segmentation accuracy of the hippocampus. In this study, we trained a novel morphological visual transformer learning-based hippocampus substructure segmentation for accurate segmentation of the anterior and posterior regions of the hippocampus from T1 weighted (T1w) MR images.
	
	\noindent 
	\section{Methods and Materials}
	\noindent 
	\subsection{Overview}
	
	Figure 1 outlines the schematic flow chart of this hippocampus multi-substructure segmentation process. The proposed network follows the same feedforward path for both training and inference. A collection of hippocampus images and multi-substructure contours was used for model training. The proposed model, named as morphological visual transformer-based network, takes the hippocampus image as input and generates the auto-contour of two substructures, which are the hippocampus proper and parts of the subiculum. The manual contours of these two substructures were used as ground truth to supervise the proposed network.
	
	The proposed model consists of two deep learning-based subnetworks, i.e., a localization model and a segmentation model. The localization model is a hippocampus ) detection network that is used to detect the volume-of-interest (VOI) for both the hippocampus proper and parts of the subiculum\cite{RN11} from the T1w MR image. The MR image is then cropped within the VOI before transfer to the segmentation subnetwork to ease the computational task. The segmentation model is implemented via an end-to-end morphological vision transformer network, which is used to perform substructures segmentation within the hippocampus VOI. The vision transformer incorporates the dominant features extracted from MR images. The integration of the morphological operators into the vision transformer increases the ability of separating the hippocampus into two substructures.
	
	During inference, the trained localization model takes a hippocampus T1w MR image as input and detects the VOI of hippocampus as the first step. Then, the cropped image within the VOI is sent to the segmentation model, i.e., morphological visual transformer, to segment the substructures. Finally, based on the detected coordinates derived by the localization model, the segmented contour is converted back to its original coordinates to obtain the final segmentation.
	
	\begin{figure}
		\centering
		\noindent \includegraphics*[width=6.50in, height=4.20in, keepaspectratio=true]{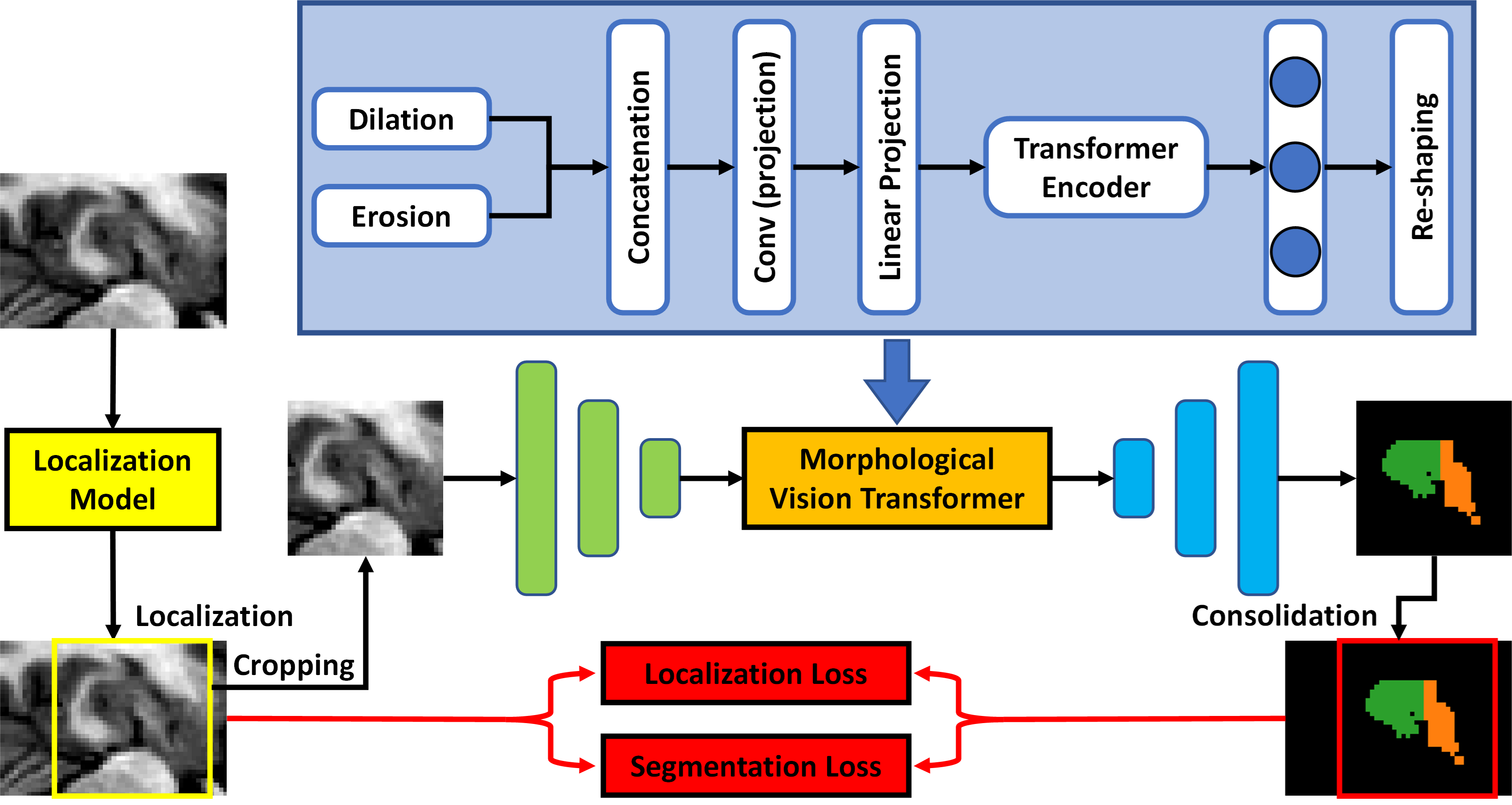}
		
		\noindent Figure 1. The workflow of the proposed morphological visual transformer learning-based hippocampus substructure segmentation.
	\end{figure}
	
	\noindent 
	\subsection{Localization model}
	
	The aim of the localization model is to crop the image to a VOI that only covers the hippocampus to ease computational task of substructure segmentation. In order to preserve the spatial information of substructure, the coordinate the detected VOI is recorded during testing. Thus, the localization of ground truth hippocampus is used to supervise the localization model. To derive it, the manual contour is needed. For a set of MR images $I_{Img} \in R^{(w\times h\times d)}$, where w and h denote the width and height of the $I_{Img}$, d represents its depth, and the corresponding physician-delineated hippocampus, $I_{Seg}=I_{seg}^p \cup I_{seg}^s I_{seg}^p$ denotes the hippocampus proper. $I_{seg}^s$ denotes the parts of the subiculum. Based on the $I_{Seg}$, the bounding box that only covers the hippocampus can be derived. This bounding box is defined as the ground truth volumes-of-interest (VOI). The coordinate of the VOI is represented by $C=[x_c,y_c,z_c,w_c,h_c,d_c ]\in R^6$, where $x_c$, $y_c$ and $z_c$ denote the center of hippocampus VOI, $w_c$, $h_c$ and $d_c$ denote the width, height and depth of the VOI along the 3D direction.
	
	The localization model design is inspired by a recently developed focal modulation network, which is used in object detection.\cite{RN12} The localization model includes a hierarchical contextualization, which is used for feature extraction from different hierarchical levels, a modulator, which combines the features from different levels, and a neural network layer works for location position estimation. The details of the localization model is explained as follows.

	Given input MRI $I_{Img} \in R^{(w\times h\times d)}$, with a first convolution layer for feature map initiating $F_0$, a multi-scale hierarchy feature map set are collected via the steps defined as follows iteratively:

	\begin{equation} 
		F_k=GeLu(Conv(F_{k-1} )),
	\end{equation} 

	where $F_{k-1}$ denotes the feature map from previous iteration, $F_k$ is then derived by the operating convolution and Gaussian error linear units (GeLU) activation function.\cite{RN19} After several iterations of Eq. (1), multi-hierarchical features are collected, we then match these feature maps to same size via interpolation and sum together
	
	\begin{equation} 
		F_m=\Sigma_k BicubicInterpolate(F_k).
	\end{equation} 
	
	Then, by using a neural network layer, we aim to derive the estimation of C, labeled as $\hat{C}=[\hat{x}_c,\hat{y}_c,\hat{z}_c,\hat{w}_c,\hat{h}_c,\hat{d}_c]$, from the $F_m$. To achieve this aim, we set the loss function, as shown in Eq. (3) during the training of localization module.
	
	\begin{equation} 
		L_{loc}=d((x_c,y_c,z_c ),(\hat{x}_c,\hat{y}_c,\hat{z}_c))+\lambda(\sqrt{w_c^2-\hat{w}_c^2}/w+\sqrt{h_c^2-\hat{h}_c^2}/h+\sqrt{d_c^2-\hat{d}_c^2}/d),
	\end{equation} 

	where $d((x_c,y_c,z_c ),(\hat{x}_c,\hat{y}_c,\hat{z}_c))$ denotes the Euclidean distance between the two centers $(x_c,y_c,z_c )$ and $(\hat{x}_c,\hat{y}_c,\hat{z}_c)$. 
	
	\noindent 
	\subsection{Morphological visual transformer}
	
	For the next step, the MRI $I_{Img}$ are cropped within a VOI box, whose center is defined as $\hat{C}$ This process mitigates the unrelated region for hippocampus segmentation and thus improve the efficiency of the model. To ensure the cropped image is uniformly sized for the following subnetwork, zero-padding is used. The processed image is then input into the morphological visual transformer (MVT). The MVT is built in an end-to-end fashion, meaning that the input and output share the same size. After several convolutional layers with a stride size of 2, the MVT uses two auto-learned morphological operators, dilation and erosion, to process the hidden feature maps. As compared to convolutional kernel with stride size of 2 or max-pooling layer, which can be regarded as a dilation with a flat square structuring element followed by a pooling, the learned morphological operator can be tuned to aggregate the most important information. This can further reduce the redundant and meaningless information for the next operator, the visual transformer, and therefore improve  its performance. The output of the two morphological operators is then concatenated and fed  into a projection convolutional layer and a linear projection operator to fit it to the input of visual transformer. A widely developed visual transformer is used.\cite{RN1} Afterwards, several deconvolutional layers are applied until the output of this MVT model  is equal in size to the input.
	
	After the MVT step, consolidation can be used to transform the segmentation back to the original coordinate system ($I_{img}$), since the location information has been obtained from the localization model.
	
	To supervise the MVT, a combination of two loss functions is used, which are generalized cross entropy loss $L_{GCE}$ and generalized Dice loss $L_{GD}$. The $L_{GCE}$ is used to evaluate the difference between the predicted label and the ground truth label at each voxel, which is defined as:
	
	\begin{equation} 
		L_{GCE}=-\Sigma_il_i\log{\hat{l}_i}
	\end{equation}
	
	where $l_i$ denotes the ground truth label at voxel i, $\hat{l}_i$ denotes the predicted label at voxel i.
	
	The $L_{GD}$ is used to address the issues about the voxel quantity imbalance of the segmented voxels (often a small portion of the whole image) and background (large portion), which is defined as:
	
	\begin{equation} 
		L_{GD}=1-2\frac{\Sigma_il_i\times\hat{l}_i+\epsilon}{\Sigma_il_i^2+\Sigma_i\hat{l}_i^2+\epsilon}
	\end{equation}

	where $\epsilon$ is a small value. The weighted sum of these two loss terms is then used to train the MVT model.
	
	\noindent 
	\subsection{Dataset}
	
	In total, 260 T1w MR images from Medical Segmentation Decathlon were used in this study.\cite{RN4} The Medical Segmentation Decathlon is a dataset consisting of T1-weighted magnetization-prepared rapid gradient echo (MPRAGE) MRIs of both healthy adults (ninety healthy adults) and adults with a non-affective psychotic disorder. The corresponding target Region of Interest (ROIs) were the anterior and posterior of the hippocampus, defined as the hippocampus proper and parts of the subiculum. This dataset was selected due to the precision needed to segment such a small object in the presence of a complex surrounding environment. 
	
	We conducted a five-fold cross-validation study on the first 200 T1w MR images. Then, a hold-out test was performed on the remaining 60 images using a model trained on the first 200 images. The segmentation was evaluated with multiple quantitative metrics including the Dice similarity coefficient (DSC), 95th percentile Hausdorff distance (HD95), mean surface distance (MSD), volume difference (VD) and center-of-mass distance (COMD). A Bland-Altman analysis and volumetric Pearson correlation analysis were also performed.

	\noindent 
	\subsection{Implementation and evaluation}
	
	The investigated deep learning networks were designed using Python 3.6 and TensorFlow and implemented on a GeForce RTX 2080 GPU that had 12GB of memory. Optimization was performed using the Adam gradient optimizer. The learning rate was 2×10-4. With the batch size setting of 20 during training, the percentage of utility of GPU memory is 96\%. Once the network was trained, it only takes ~1.5 mins for hippocampus segmentation. To demonstrate the utility of morphological operator, an ablation study was conducted. Namely, we tested the performance of the proposed method of with and without using morphological operator. To further demonstrate the significance of the proposed work, we compared the proposed method with another popular segmentation models, cascaded U-Net (CasU)\cite{RN2} and visual transformer network (VIT).\cite{RN1} Comparisons were performed using the same training and testing datasets and computational environment.

	\noindent 
	\section{Results}
	\noindent 
	\subsection{Comparing with state-of-the-art}
	
	The visual comparison between the proposed method and comparing methods are shown in Fig. 2. As can be seen from the first row, the proposed method shows good agreement with the ground truth, whereas the comparing methods cannot. In the second row it is observed that misclassification of posterior part occurs for the cascaded U-Net. To better demonstrate the segmentation accuracy, we performed absolute subtraction of the segmentation results of the proposed method and comparing methods with the manual contour’s binary masks. The difference images are shown in the fourth to sixth rows. As can be seen from the fifth and sixth rows, the difference images of the two comparing methods show greater error at the adjacent part between the hippocampus proper and parts of the subiculum.
	
	\begin{figure}
		\centering
		\noindent \includegraphics*[width=6.50in, height=4.20in, keepaspectratio=true]{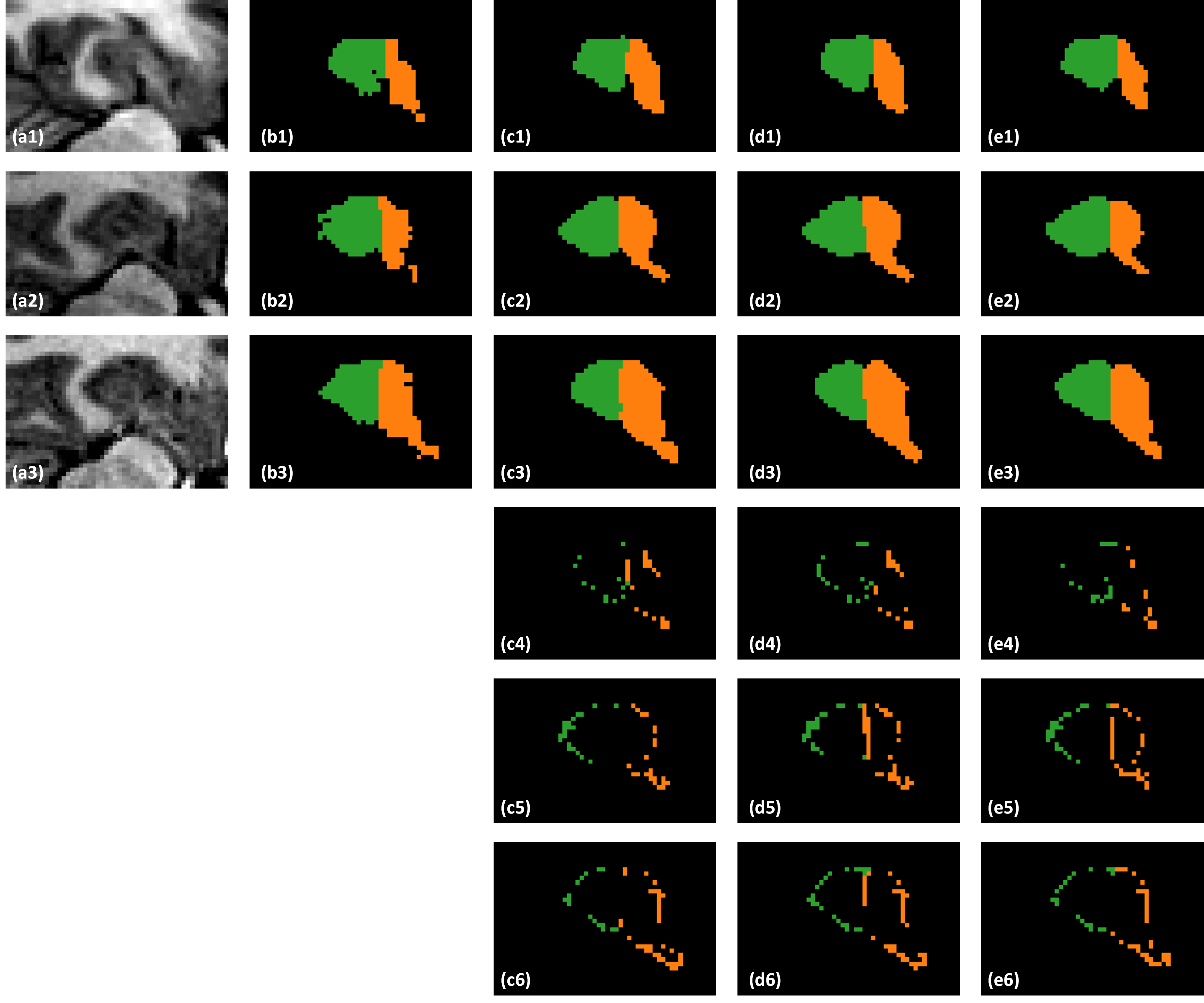}
		
		\noindent Figure 2. A representative case of proposed method and state-of-the-art methods. The 1st column shows MR images. The 2nd column shows the ground truth contour. The 3rd column shows the results of proposed method. The 4th column and 5th column show the results of cascaded U-Net and VIT, respectively. The last three rows are related to the absolute difference between segmented ones and ground truth ones.
	\end{figure}

	The linear correlation coefficient calculated as target volume of ground truth and segmentation, is shown in Fig. 3. The linear correlation coefficient obtained using the proposed method was 0.999 and 0.993 on five-fold cross-validation and hold-out test, respectively. These values indicate a good agreement between the ground truth and proposed results, as compared to 0.989/0.983 and 0.991/0.979 obtained by the cascaded U-Net and VIT, respectively on five-fold cross-validation/hold-out test. On hold-out test, the VIT consistently underestimated the region , which became more pronounced for larger tumors.

	\begin{figure}
		\centering
		\noindent \includegraphics*[width=6.50in, height=4.20in, keepaspectratio=true]{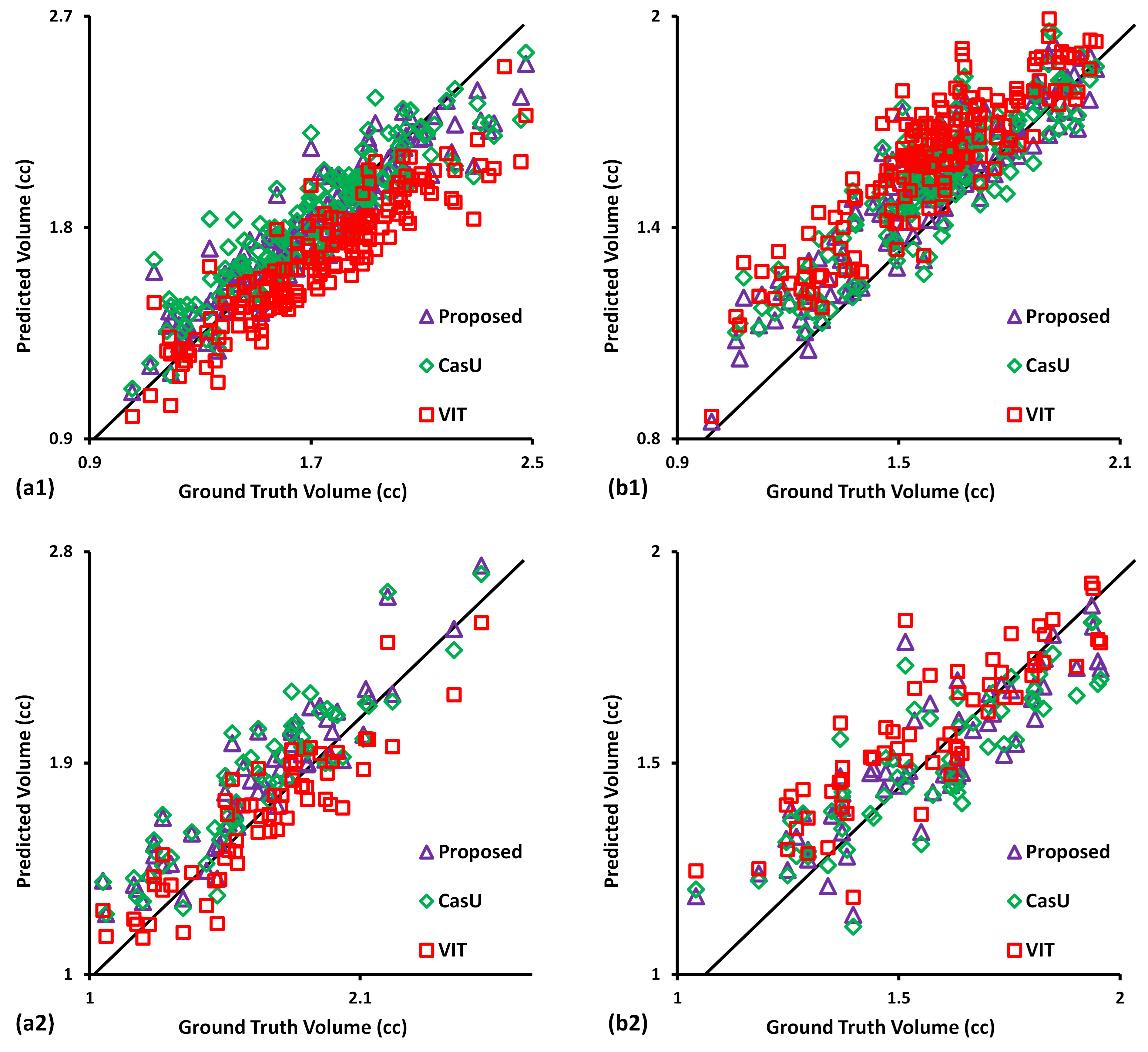}
		
		\noindent Figure 3. Bland-Altman analysis of the segmented volumes between ground truth (semi-log scale) against the proposed method and comparing methods. Each dot indicates a data point from the dataset for that model. (a) row denotes the results of five-fold cross-validation. (b) row denotes the results of hold-out test. First column denotes the segmentation of first substructure. Second column denotes the segmenting results of second substructure.
	\end{figure}
	
	The quantitative metrics of the proposed method and the alternate methods from the 200 cases’ cross-validation and 60 cases’ hold-out test are listed in Table 1 and 2, and Table 3 and 4, respectively. For the cross-validation experiment, the proposed model significantly outperformed Cascaded U-Net and VIT in all metrics. In five-fold cross-validation, the DSCs, HD95, MSD and CMD were 0.900±0.029 and 0.886±0.031, 1.156±0.277 and 1.133±0.264, 0.426±0.115 and 0.401±0.100, 0.491±0.300 and 0.738±0.452 for the hippocampus proper and parts of the subiculum, respectively.
	
	\begin{table}[]
		\centering
		\caption{Numerical results (hippocampus proper) on 5-fold cross-validation of proposed method, cascaded U-Net and VIT, respectively.}
		\label{tab:my-table}
		\resizebox{\textwidth}{!}{%
			\begin{tabular}{lllllll}
				\hline
				& DSC & Jac & HD95 (mm) & MSD (mm) & RMSD (mm) & CMD (mm) \\ \hline
				Proposed & 0.900±0.029 & 0.819±0.047 & 1.156±0.277 & 0.426±0.115 & 0.688±0.122 & 0.491±0.300 \\ \hline
				CasU & 0.891±0.031 & 0.804±0.049 & 1.329±0.478 & 0.466±0.123 & 0.744±0.156 & 0.581±0.363 \\ \hline
				VIT & 0.894±0.027 & 0.809±0.044 & 1.195±0.296 & 0.441±0.107 & 0.706±0.112 & 0.573±0.297 \\ \hline
			\end{tabular}%
		}
	\end{table}
	
	\begin{table}[]
		\centering
		\caption{Numerical results (parts of the subiculum) on 5-fold cross-validation of proposed method, cascaded U-Net and VIT, respectively.}
		\label{tab:my-table}
		\resizebox{\textwidth}{!}{%
			\begin{tabular}{lllllll}
				\hline
				& DSC & Jac & HD95 (mm) & MSD (mm) & RMSD (mm) & CMD (mm) \\ \hline
				Proposed & 0.886±0.031 & 0.796±0.049 & 1.133±0.264 & 0.401±0.100 & 0.677±0.109 & 0.738±0.452 \\ \hline
				CasU & 0.874±0.033 & 0.778±0.051 & 1.291±0.415 & 0.443±0.111 & 0.735±0.137 & 0.948±0.597 \\ \hline
				VIT & 0.882±0.030 & 0.791±0.047 & 1.215±0.324 & 0.42±0.100 & 0.707±0.115 & 0.798±0.509 \\ \hline
			\end{tabular}%
		}
	\end{table}

	In the hold-out test using external datasetthe proposed model is significantly superior to the alternate approaches, as shown in Table 3 and 4 in comparison with cascaded U-Net and VIT. In hold-out test, the DSCs, HD95, MSD and CMD were 0.881±0.033 and 0.863±0.034, 1.328±0.404 and 1.272±0.388, 0.494±0.113 and 0.466±0.112, 0.608±0.313 and 0.834±0.478 for the hippocampus proper and parts of the subiculum, respectively. As compared to five-fold cross-validation, the hold-out test did slightly worse with slightly higher standard deviation, which may be caused by the training data’s distribution not covering the range of cases in the hold-out test.
	
	\begin{table}[]
		\centering
		\caption{Numerical results (hippocampus proper) on hold-out test of proposed method, cascaded U-Net and VIT, respectively.}
		\label{tab:my-table}
		\resizebox{\textwidth}{!}{%
			\begin{tabular}{lllllll}
				\hline
				& DSC & Jac & HD95 (mm) & MSD (mm) & RMSD (mm) & CMD (mm) \\ \hline
				Proposed & 0.881±0.033 & 0.789±0.052 & 1.328±0.404 & 0.494±0.113 & 0.754±0.131 & 0.608±0.313 \\ \hline
				CasU & 0.871±0.032 & 0.773±0.051 & 1.478±0.537 & 0.535±0.118 & 0.81±0.155 & 0.703±0.35 \\ \hline
				VIT & 0.876±0.030 & 0.781±0.047 & 1.378±0.455 & 0.501±0.100 & 0.774±0.132 & 0.683±0.336 \\ \hline
			\end{tabular}%
		}
	\end{table}

	\begin{table}[]
		\centering
		\caption{Numerical results (parts of the subiculum) on hold-out test of proposed method, cascaded U-Net and VIT, respectively.}
		\label{tab:my-table}
		\resizebox{\textwidth}{!}{%
			\begin{tabular}{lllllll}
				\hline
				& DSC & Jac & HD95 (mm) & MSD (mm) & RMSD (mm) & CMD (mm) \\ \hline
				Proposed & 0.863±0.034 & 0.761±0.051 & 1.272±0.388 & 0.466±0.112 & 0.742±0.126 & 0.834±0.478 \\ \hline
				CasU & 0.852±0.036 & 0.744±0.053 & 1.419±0.565 & 0.509±0.123 & 0.801±0.165 & 0.986±0.6 \\ \hline
				VIT & 0.858±0.035 & 0.753±0.053 & 1.349±0.435 & 0.491±0.115 & 0.776±0.141 & 0.893±0.572 \\ \hline
			\end{tabular}%
		}
	\end{table}

	\bigbreak
	
	\noindent 
	\section{Discussion}
	
	A novel hippocampus segmentation method (called MVT) is proposed by introducing a localization mechanism to aid segmentation and designing the morphological visual transformer network for substructures segmentation. The localization model is used to detect the VOI of hippocampus. The end-to-end morphological vision transformer network is used to perform substructures segmentation within the hippocampus VOI. The substructures include the anterior and posterior regions of the hippocampus, which are defined as the hippocampus proper and parts of the subiculum. The vision transformer incorporates the dominant features extracted from MRI images and is improved by learning-based morphological operators. The morphological operators integrated into the vision transformer enhance the ability to separate the hippocampus structure into two substructures. 
	
	Due to limited computational resources, our method focused on domain incremental learning with a cropped region for analysis. We plan to test the performance of our method in a class incremental setup. As the visual transformer contains several orders of magnitude larger number of parameters due to the self-adapting process as compared to the traditional CNNs, it is essential to investigate an effective optimization method to reduce the amount of GPU memory allocation as well as simplify the overall ViT U-Net architecture.
	
	Our MVT is a supervised method, which means it still requires accurate manual contours as training labels. Currently, there are semi-supervised learning methods that can learn features from unlabeled data. We will extend the proposed method with the ensemble approach to improve its generalization performance by integrating the supervision learning and semi-supervised learning methods from the limited labeled data and large-scale unlabeled data of MRIs in a future study.
	
	The auto-segmentation of substructures of hippocampus has significant  clinical relevance. For example, in hippocampal sparing whole brain radiation therapy (HA-WBRT),\cite{RN24} current intensity modulated radiation treatment (IMRT) and arc-based VMAT techniques can reduce dose to the hippocampus without sacrificing target coverage and homogeneity.\cite{RN25} Further improvements in patient outcomes may be possible by considering substructures separately for optimal dose sparing; however, accurate segmentation is critical. With more accurate contouring of substructures of hippocampus, it is possible to have different dose constraints of these substructures in HA-WBRT,\cite{RN26} allowing for better sparing of the critical part of the hippocampus.

	\bigbreak
	
	\noindent 
	\section{Conclusion}
	
	We have developed a novel deep learning-based method to accurately segment the anterior and posterior of hippocampus. Our results showed good performance in terms of DSC and VD between the segmentation result and the ground truth.

	\noindent 
	\bigbreak
	{\bf ACKNOWLEDGEMENT}
	
	This research is supported in part by the National Institutes of Health under Award Number R01CA215718, R56EB033332, R01EB032680, and P30CA008748.

	\noindent 
	\bigbreak
	{\bf Disclosures}
	
	The authors declare no conflicts of interest.

	\noindent 
	
	\bibliographystyle{plainnat}  
	\bibliography{arxiv}      
	
\end{document}